\documentclass[reprint,
amsmath,amssymb,
aps,
]{revtex4-2}
\usepackage{graphicx}
\usepackage{dcolumn}
\usepackage{bm}
\usepackage{url}
\usepackage{here}
\usepackage{tabularx}

\begin{document}

\title{Thermodynamic Properties of the Mott Insulator-Metal Transition\\
 in a Triangular Lattice System Without Magnetic Order}

\author{Emre Yesil$^{1}$}
\author{Shusaku Imajo$^{2}$}
\email{imajo@issp.u-tokyo.ac.jp}
\author{Satoshi Yamashita$^{1}$}
\author{Hiroki Akutsu$^{1}$}
\author{Yohei Saito$^{3}$}
\author{Andrej Pustogow$^{4}$}
\author{Atsushi Kawamoto$^{5}$}
\author{Yasuhiro Nakazawa$^{1}$}
\email{nakazawa@chem.sci.osaka-u.ac.jp}
\affiliation{
$^1$Graduate School of Science, Osaka University, Toyonaka, Osaka 560-0043, Japan\\
$^2$Institute for Solid State Physics, University of Tokyo, Kashiwa, Chiba 277-8581, Japan\\
$^3$Institute of Physics, Goethe-University Frankfurt, 60438 Frankfurt (M), Germany\\
$^4$Institute of Solid State Physics, TU Wien, 1040 Vienna, Austria\\
$^5$Graduate School of Science, Hokkaido University, Sapporo 060-0810, Japan
} 
\date{\today}
\begin{abstract}
The organic system, $\kappa$-[(BEDT-TTF)$_{1-x}$(BEDT-STF)$_x$]$_2$Cu$_2$(CN)$_3$, showing the Mott transition between a nonmagnetic Mott insulating (NMI) state and a Fermi liquid (FL), is systematically studied by calorimetric measurements.
An increase of the electronic heat capacity at the transition from the NMI state to the FL state which keeps the triangular dimer lattice demonstrates that the charge sector lost in the Mott insulating state is recovered in the FL state.
We observed that the remaining low-energy spin excitations in the Mott insulating state show unique temperature dependence, and that the NMI state has a larger lattice entropy originating from the frustrated lattice, which leads to the Pomeranchuk-like effect on the electron localization.
Near the Mott boundary, an unexpected enhancement and magnetic-field dependence of heat capacity are observed.
This anomalous heat capacity is different from the behavior in the typical first-order Mott transition and shows similarities with quantum critical behavior.
To reconcile our results with previously reported scenarios about a spin gap and the first-order Mott transition, further studies are desired.
\end{abstract}

\maketitle
\section{Introduction}
The dimer-Mott compounds with the chemical formula of $\kappa$-(BEDT-TTF)$_2$$X$, where BEDT-TTF is bis(ethylenedithio)tetrathiafulvalene and $X$ is a monovalent counter anion, provide extensive possibilities for understanding physical phenomena induced by electron correlations of $\pi$-electrons\cite{1,2}.
%---------------------
\begin{figure*}
\begin{center}
\includegraphics[width=1\linewidth,clip]{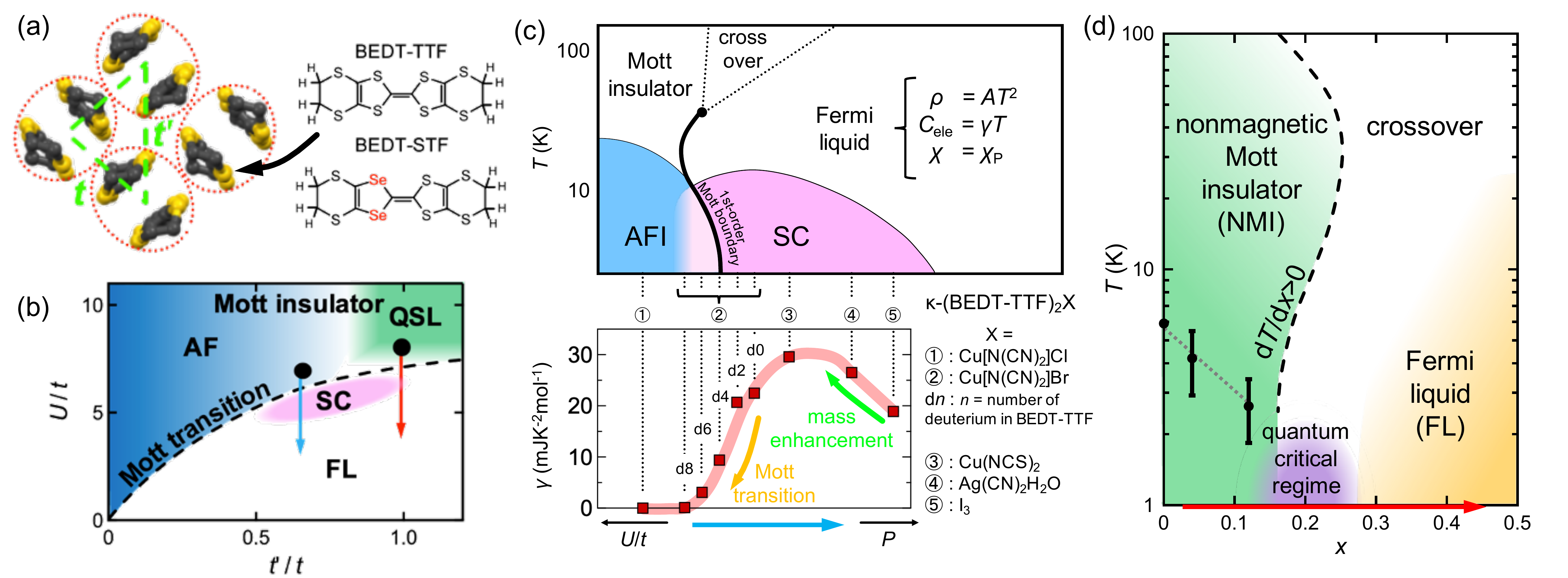}
\end{center}
\caption{(Color online)
(a) Molecular arrangement in the conducting plane of the present system and BEDT-TTF and BEDT-STF molecules.
$t$ and $t$$^{\prime}$ indicate nearest-neighbor and second-nearest-neighbor transfer integrals in the dimer lattice, respectively.
(b) Electronic ground states of dimer-Mott system with parameters of electron correlations $U$/$t$ ($U$/$W$) and frustration factor $t$$^{\prime}$/$t$\cite{6,7}.
The red arrow indicates the route controlled by substitution in the present system, whereas the light blue arrow represents the numerous previous studies on the less-frustrated dimer-Mott system.
(c) Electronic phase diagram of the less-frustrated $\kappa$-type salts with the shown counter anions. 
The AFI and FL phases are divided by the first-order Mott transition, which terminates at a critical endpoint $\sim$35~K.
Inside the FL, SC with relatively high-$T$$_{\rm c}$$\sim$10~K occurs near the boundary.
The lower panel shows the variation in $\gamma$ of the normal state depending on electron correlations $U$/$t$ and chemical pressure of the counter anion $P$.
(d) Schematic phase diagram of $\kappa$-[(BEDT-TTF)$_{1-x}$(BEDT-STF)$_x$]$_2$Cu$_2$(CN)$_3$ deduced from Refs.~\cite{26,27,30,31}.
An increase in the mixing ratio $x$ corresponds to a decrease in $U$/$t$ shown by the red arrow in the phase diagram (b).
The NMI and electronic FL states exist across the quantum critical regime, which is located around $x$=0.2.
The dashed curve indicates the phase boundary between the NMI and FL phases, its slope d$T$/d$x$ is positive at low temperature.
The black dots in the NMI region represent the $x$ dependence of the so-called 6~K anomaly, in which error bars are determined by the present heat capacity measurements.
}
\label{fig1}
\end{figure*}
%---------------------
Electrons in them form relatively narrow electron bands governed by overlaps of molecular orbitals, and the spin, charge, and lattice degrees of freedom appear in various manners in them.
The electronic states of the dimer-Mott system can be described in the frame of the Mott-Hubbard physics with on-site Coulomb repulsion $U$ and bandwidth $W$ (proportional to transfer integral $t$)\cite{1,2,3,4,5}.
Additionally, the dimer lattice of the $\kappa$-type molecular arrangement has geometrical frustration depending on the ratio of $t$ and $t$$^{\prime}$, nearest-neighbor and second-nearest-neighbor transfer integrals, as shown in Fig.~\ref{fig1}(a).
Using the two parameters $U$/$t$ vs. $t$$^{\prime}$/$t$, the electronic phase diagram has been understood, as shown in Fig.~\ref{fig1}(b)\cite{6,7}.
For the less-frustrated salts ($t$$^{\prime}$/$t$$<$1), the discontinuity and hysteresis in the electrical transport indicate that the superconductivity-antiferromagnetic insulator (SC-AFI) transition dominated by a change in $U$/$t$ (the blue arrow in Fig.~\ref{fig1}(b)) is first-order\cite{8,9,10,11,12}.
As schematically described in Fig.~\ref{fig1}(c), the 1st-order Mott boundary disappears at a critical endpoint of $\sim$35~K, and the nature of the Mott physics around the endpoint has been discussed in terms of high-energy criticality caused by the competition between the large $U$ and $W$ $>$1000~K\cite{10,11,12}.
From the AF magnetic order induced by antiferromagnetic interactions, the SC with relatively high-$T$$_{\rm c}$ has been extensively discussed in terms of unconventional pairing related to antiferromagnetic spin fluctuations in $\kappa$-(BEDT-TTF)$_2$$X$ and also in $\beta$$^{\prime}$-, $\lambda$-type compounds\cite{13,13p5,14,14p5,15}.
The variation in physical parameters near the Mott boundary has been studied by various measurements across the boundary\cite{16,17,18,19,20}. 
Based on the variation in the electronic heat capacity coefficient $\gamma$ of the normal state shown in Fig.~\ref{fig1}(c)\cite{13p5,36,36p1,36p2,36p3}, the low-temperature FL state can be understood by the electron-mass enhancement with increasing electron correlations (the green arrow) and the decrease in the metallic portion due to the growth of phase coexistence near the Mott boundary (the orange arrow).
Although slight percolative superconductivity is left in the AF Mott insulating salts very near the boundary, its $\gamma$ is almost zero because the volume fraction of the FL is negligible.
It should be noted that the information contains the magnetic entropy change related to the AFI ground state of $\pi$-electrons and that the change is not a genuine feature expected in the Hubbard model because no symmetry breaking is assumed in this framework\cite{21,22,23,24,25}.
When $t$$^{\prime}$/$t$=1, the AFI ground state should be destabilized by the geometrical frustration, and non-ordered states may be stable even at low temperatures.
Indeed, $\kappa$-(BEDT-TTF)$_2$Cu$_2$(CN)$_3$, which has been considered a prime candidate showing the quantum spin liquid (QSL) state, does not show long-range magnetic orders down to extremely low temperatures because $t$$^{\prime}$/$t$ is almost unity\cite{28,31p2}.
However, recently, Miksch et al. suggested that its ground state might be a gapped valence bond solid (VBS) with a spin gap from observation of a drop of spin susceptibility below 6~K\cite{25,35}.
Although the controversy still persists because of the remaining discrepancy with the gapless spin excitations in heat capacity\cite{29}, we hereafter use NMI for describing the Mott insulating state.

Recently, Saito et al. reported that a donor alloying system of (BEDT-TTF)$_{1-x}$(BEDT-STF)$_x$, where BEDT-TTF and BEDT-STF are the abbreviations of bis(ethylenedithio)tetrathiafulvalene and bis(ethylenedithio)diselenadithiafulvalene, with Cu$_2$(CN)$_3$$^-$ exhibits continuous tuning of $U$/$W$ with keeping the triangularity of the dimer lattice\cite{26,27,30,31}.
The Se substitution into the BEDT-TTF molecule shown in Fig.~\ref{fig1}(a) results in a larger overlap of the wave function with the neighboring molecules.
Since the increase in $x$ is considered to work as positive chemical pressure without inducing large change in average $t$$^{\prime}$/$t$, the insulating state is altered into the FL state via the genuine Mott transition at $x$=0.1-0.2\cite{26,27,30,31}, as indicated by the red arrow in Fig.~\ref{fig1}(b).
This variation is similar to the tuning by external pressure to $\kappa$-(BEDT-TTF)$_2$Cu$_2$(CN)$_3$ where the ground state without magnetic order shifts to a FL across the Mott insulator-metal transition\cite{8,9}.
Namely, this variation provides profound information on the Mott transition genuinely dominated by the itinerancy/localization of the charge degrees of freedom.
The $T$-$x$ electronic phase diagram of the present alloying system is predicted from the results in Refs.~\cite{26,27,30,31}, as shown in Fig.~\ref{fig1}(d).
High-resolution thermodynamic measurements under pressure are typically challenging; however, using the present chemically tunable system, thermodynamic and entropic information near the metal-insulator boundary can be obtained by ambient-pressure heat capacity measurements.
In this study, we systematically investigated $\kappa$-[(BEDT-TTF)$_{1-x}$(BEDT-STF)$_x$]$_2$Cu$_2$(CN)$_3$ by calorimetry to unveil thermodynamics features of the Mott transition between the potential QSL and FL states.
%%%%%%%%%%
\section{Experimental}
Single crystals of the alloying compounds $\kappa$-[(BEDT-TTF)$_{1-x}$(BEDT-STF)$_x$]$_2$Cu$_2$(CN)$_3$ are grown by electrochemical oxidation method\cite{27}.
As shown in Table~\ref{tab1}, the crystal structural parameters were characterized by x-ray diffraction analyses, and the macroscopic homogeneity of the alloying crystals was confirmed.
To evaluate the change in $t$$^{\prime}$/$t$ with mixing BEDT-STF molecules, we here introduce $d$$^{\prime}$/$d$, the ratio of average dimer-dimer distance along the $t$$^{\prime}$ and $t$ directions.
The small changes in the lattice parameters within 0.5$\%$ were observed.
Heat capacity measurements were carried out by a typical relaxation technique using a home-made thermal-relaxation-type calorimeter in a $^{3}$He refrigerator with a 15~T superconducting magnet.
The temperature range of these measurements is about 0.6-10 K.
Magnetic fields were applied perpendicular to the conducting plane.
We measured the background data with a small amount of Apiezon N grease before mounting samples.
These measurements were performed with single crystalline samples weighing about 80-300~$\mu$g.
The details of the calorimeter and experimental setup are reported in Ref.~\cite{31p5}.
\begin{table}
\caption{Crystallographic data.
$Fw$ represents the formula weight for each sample.
$V$ shows the cell volume.
$Z$ denotes the number of formula units in the unit cell divided by the number of independent general positions.
$d$$^{\prime}$/$d$ is the ratio of average dimer-dimer distance along the $t$$^{\prime}$ and $t$ directions.}
\label{tab1}
\begin{center}
\scalebox{0.9}{
\begin{tabular}{ccccccc}
\hline \hline
$x$ & 0.04  & 0.10  & 0.12  &  0.19  &  0.28  &  0.44\\
\hline
$Fw$  & 982.04  & 993.30 & 997.05  & 1010.18  & 1027.06  & 1057.07\\
Space group  & $P2_{1}/c$  & $P2_{1}/c$  & $P2_{1}/c$  & $P2_{1}/c$  & $P2_{1}/c$  & $P2_{1}/c$ \\
$a$ (${\rm \AA}$)  &  16.1080  &  16.1054  &  16.1136  &  16.1569  &  16.1580  &  16.1761\\
$b$ (${\rm \AA}$)  &  8.5861  &  8.5816  &  8.5874  &  8.6000  &  8.6017  &  8.5982\\
$c$ (${\rm \AA}$)  &  13.3591  &  13.3751  &  13.3550  &  13.3663  &  13.3979  &  13.4037\\
$\alpha$ ($^{\circ}$)  &  90  &  90  &  90  &  90  &  90  &  90\\
$\beta$ ($^{\circ}$)  &  113.691  &  113.565  &  113.66  &  113.519  &  113.551  &  113.208\\
$\gamma$ ($^{\circ}$)  &  90  &  90  &  90  &  90  &  90  &  90\\
$V$ (${\rm \AA}$$^3$)  &  1691.9  &  1694.4  &  1692.6  &  1703.0  &  1707.0 &  1713.4\\
$Z$  & 2  & 2  & 2  & 2  & 2  & 2 \\
$d$$^{\prime}$/$d$  & 0.925  & 0.926  & 0.925  & 0.924  & 0.925  & 0.926 \\
\hline
\end{tabular}
}
\end{center}
\end{table}

\section{Results}
\begin{figure*}
\begin{center}
\includegraphics[width=1\linewidth,clip]{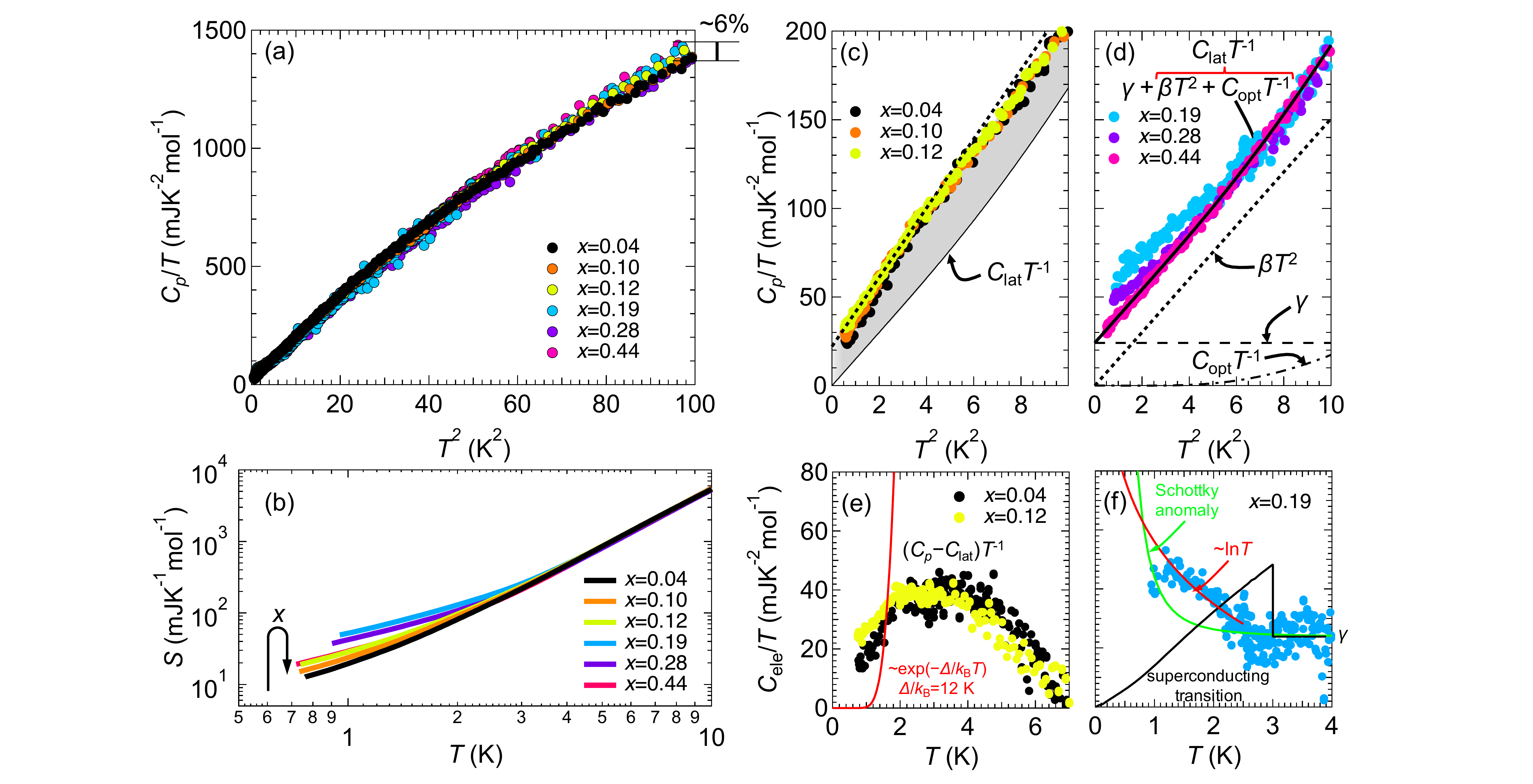}
\end{center}
\caption{(Color online)
(a) $C_p$$T$$^{-1}$ vs. $T$$^2$ below 10~K for $x$=0.04-0.44.
(b) Logarithmis plot of $S$ as a function of temperature.
(c),(d) Enlarged plots of the low-temperature region below $T$$<$3.2~K for $x$$<$0.15 (c) and $x$$>$0.15 (d).
The dotted line in (c) is a fit to $C_p$$T$$^{-1}$=$\gamma$+$\beta$$T^2$ below 2~K for $x$=0.12.
The thin black curve in (c) is a rough estimate of lattice heat capacity $C$$_{\rm lat}$$T$$^{-1}$ obtained from the FL salt ($x$=0.44) by subtracting the $\gamma$ term, which highlights the contribution of the low-energy excitations in the NMI state (shaded area).
The black curves in (d) show the respective components of the fit of $C_p$$T$$^{-1}$=$\gamma$+$\beta$$T^2$+$C$$_{\rm opt}$$T$$^{-1}$ to the $x$=0.44 data.
(e),(f) Electronic heat capacity $C_{\rm ele}$$T^{-1}$ obtained by subtracting the lattice heat capacity $C$$_{\rm lat}$$T$$^{-1}$ from the total $C_p$$T$$^{-1}$ for $x$=0.04, 0.12 (e) and 0.19 (f).
The red curve in (e) shows activation-type gapped behavior when $\Delta$/$k$$_{\rm B}$=12~K.
The red curve in (f) indicates a fit to $-$ln$T$ while the black and green curves represent the typical superconducting and Schottky anomalies, respectively.
}
\label{fig2}
\end{figure*}

In Fig.~\ref{fig2}, we present the temperature dependences of the heat capacity of the alloying system in the $C_p$$T$$^{-1}$ vs. $T^2$ plot.
%---------------------
%---------------------
The data in the temperature range up to 10~K are displayed in Fig.~\ref{fig2}(a).
There is only a subtle difference of about 6$\%$ at 10~K, mainly originating from the decrease in the Debye temperature induced by the Se substitution\cite{42p7}.
This means that the STF substitution induces only a small change of about a few percent in phonon contributions.
In Fig.~\ref{fig2}(b), the entropy $S$ as a function of temperature is shown as a logarithmic plot.
The entropy is calculated by integration of the experimentally obtained $C_p$$T$$^{-1}$ and the extrapolation down to 0~K estimated by polynomial fittings, and thus, the calculated $S$ includes the electronic and phonon contributions together.
At higher temperatures, the $x$ dependence of the entropy is small because the main portion of the total entropy is the phonon contribution.
In the lower temperature region, the $x$=0.19 salt shows the larger entropy compared to the others.
To shed light on the low-temperature region, the enlarged plots of $C_p$$T$$^{-1}$ below about 3.2~K (=10~K$^2$) are shown in Fig.~\ref{fig2}(c) and (d).
The datasets for $x$$<$0.15 are shown in Fig.~\ref{fig2}(c) while those for $x$$>$0.15 are in Fig.~\ref{fig2}(d) because the Mott insulating character at the low-$x$ region changes into the metallic one across the boundary region at $x$=0.1-0.2 according to the previous reports\cite{26,30,31}.
The small change in lattice heat capacity indicates that the origin of the change observed in the low-temperature region should mainly come from the electronic contribution.
In the case of typical metals having Fermi surfaces composed of itinerant electrons, $C_p$$T$$^{-1}$ at low temperatures obeys $C_p$$T$$^{-1}$=$\gamma$+$\beta$$T^2$, where $\gamma$ and $\beta$ represent the Sommerfeld coefficient of electronic heat capacity and the Debye coefficient of lattice heat capacity.
Indeed, the $x$=0.44 salt, which is deep inside the metallic FL region, shows the linear behavior below 2~K with $\gamma$=24.1~mJK$^{-2}$mol$^{-1}$ and $\beta$=15.0~mJK$^{-4}$mol$^{-1}$, which are comparable with those of typical BEDT-TTF-based metallic salts\cite{13p5,14p5,32}.
Above 2~K, the behavior is gradually deviated by excess heat capacity that may originate from librational optical modes $C$$_{\rm opt}$$\sim$$R$($T_{\rm E}$/$T$)exp($T_{\rm E}$/$T$)/[exp($T_{\rm E}$/$T$)$-$1]$^2$, where $T_{\rm E}$ represents the Einstein temperature, as suggested for the other organic charge-transfer complexes with various structures\cite{33}.

On the other hand, the insulating salts shown in Fig.~\ref{fig2}(d) do not share this behavior.
At first glance, it appears to follow the linear behavior below 2~K, as indicated by the black dotted line. 
Also, the analysis of the data for $x$=0.04 using the typical $C_p$$T$$^{-1}$=$\gamma$+$\beta$$T^2$ relation leads to $\gamma$=12.6~mJK$^{-2}$mol$^{-1}$ and $\beta$=21.2~mJK$^{-4}$mol$^{-1}$, which are comparable with the previously reported $\gamma$=12~mJK$^{-2}$mol$^{-1}$ and $\beta$=21~mJK$^{-4}$mol$^{-1}$ for $x$=0\cite{29}.
However, above 2~K, the $C_p$$T$$^{-1}$ is lower than this linear dependence.
Since higher-order terms of the Debye model appears only at higher temperatures, this behavior indicates that the Mott insulating state cannot be explained by the framework of the typical FL states.
Nevertheless, the large low-temperature heat capacity in the insulating state proves the presence of low-energy spin excitations, which have been discussed as the finite $\gamma$ and/or the relatively large $\beta$ specific to the organic QSL state in the previous works\cite{29,34}.
As a rough estimate, we show the lattice heat capacity $C$$_{\rm lat}$$T$$^{-1}$, which is simply obtained by subtracting the $\gamma$ term from the $x$=0.44 data (the thin black line), $C_p$$T$$^{-1}$=$\gamma$+$\beta$$T^2$+$C$$_{\rm opt}$$^{-1}$, as is shown in Fig.~\ref{fig2}(d).
The difference from this estimate ($C_p$$-$$C$$_{\rm lat}$)$T^{-1}$, which corresponds to the contribution of the spin excitations, is displayed as a $C_{\rm ele}$$T^{-1}$ vs. $T$ plot in Fig.~\ref{fig2}(e).
This component does not appear to be a simple $\gamma$ term.
ESR results\cite{35} suggest that the ground state is a gapped VBS state with a relatively large $\Delta$/$k$$_{\rm B}$$\sim$12~K.
However, the red curve, exp($-$$\Delta$/$k$$_{\rm B}$$T$) behavior for $\Delta$/$k$$_{\rm B}$=12~K, does not describe the present results.
Even if we assume that $\Delta$/$k$$_{\rm B}$ is a variable parameter, it is difficult to reproduce the temperature dependence and $\Delta$/$k$$_{\rm B}$ must be extremely tiny.
Including the present result, heat capacity measurements, sensitive to low-energy excitations, indicate the presence of low-energy spin excitations, which is puzzling in view of a spin gap concluded from other measurements\cite{35,34p5}.
To clarify this point, experiments at lower temperatures seem necessary.
Although the drop of the magnetic susceptibility below 6~K is observed, an exact zero susceptibility in a low-temperature limit has not been reported in these works\cite{35}.
To reconcile these arguments based on the temperature range of the measurements (ESR measurement above 2~K), one possibility is that the ground state has an incomplete spin gap, yielding some low-energy excitations, even below the putative transition at 6~K.
Alternatively, an extrinsic origin, such as impurity spins or domain walls, was suggested to describe the low-temperature magnetic behavior\cite{25,34p7}. 
However, it is unclear how to model the present temperature dependence with the suggested local orphan spins and local domain wall fluctuations, which may give the Schottky-type heat capacity and glass-like $\gamma$$T$ heat capacity, respectively.

For the $x$=0.19 salt, located in the intermediate region\cite{26,31}, the temperature dependence (Fig.~\ref{fig2}(d)) does not obey the $C_p$$T$$^{-1}$=$\gamma$+$\beta$$T^2$ relation due to the gradual upward deviation below $\sim$2.5~K.
This behavior is more clear in the plot of ($C_p$-$C$$_{\rm lat}$)$T^{-1}$, as shown in Fig.~\ref{fig2}(f).
Even though remnants of superconductive components are observed near the 1st-order Mott boundary of several $\kappa$-type salts, including some STF compounds\cite{42p5}, this behavior is completely distinct from the superconducting transition (the black curve).
Furthermore, this deviation cannot be reproduced by extrinsic Schottky anomaly arising from magnetic impurities (the green curve).
In the case of the AFI-FL Mott transition, such behavior is absent, and the simple $C_p$$T$$^{-1}$=$\gamma$+$\beta$$T^2$ relation is observed even very near the first-order Mott boundary\cite{36,36p1,36p3}.
The low-temperature gradual divergence is reminiscent of quantum critical behavior near a quantum critical point (QCP) because of the $-$ln$T$-like behavior (the red curve).
Indeed, the present alloying system does not show significant first-order-like discontinuous behavior in our heat capacity data and resistivity data\cite{26,30}, albeit a dielectric catastrophe suggestive of phase inhomogeneity has been reported\cite{31}.
The first-order Mott transition observed in other $\kappa$-type salts is less obvious in $\kappa$-(BEDT-TTF)$_2$Cu$_2$(CN)$_3$; nevertheless, the weak first-order Mott transition with the critical endpoint located at 15-20~K has been observed in transport, NMR, and dielectric measurements\cite{8,9,31}.
A small difference between the alloying system and $\kappa$-(BEDT-TTF)$_2$Cu$_2$(CN)$_3$ may further lead to suppression of the remaining first-order nature.
Randomness effects should be also taken into account, as a disorder can lower the temperature of the critical endpoint\cite{43}.
Regardless of the origin of the suppression of the first-order nature, the low-temperature diverging heat capacity indicates that quantum fluctuations are developed in this temperature region ($<$2.5~K).
Since this behavior is significant in $x$=0.19 and smaller in $x$=0.28, the QCP should be located close to $x$=0.2, which is not far from the reported position of the metal-insulator transition\cite{26,30,31}.

%---------------------
\begin{figure*}
\begin{center}
\includegraphics[width=1\linewidth,clip]{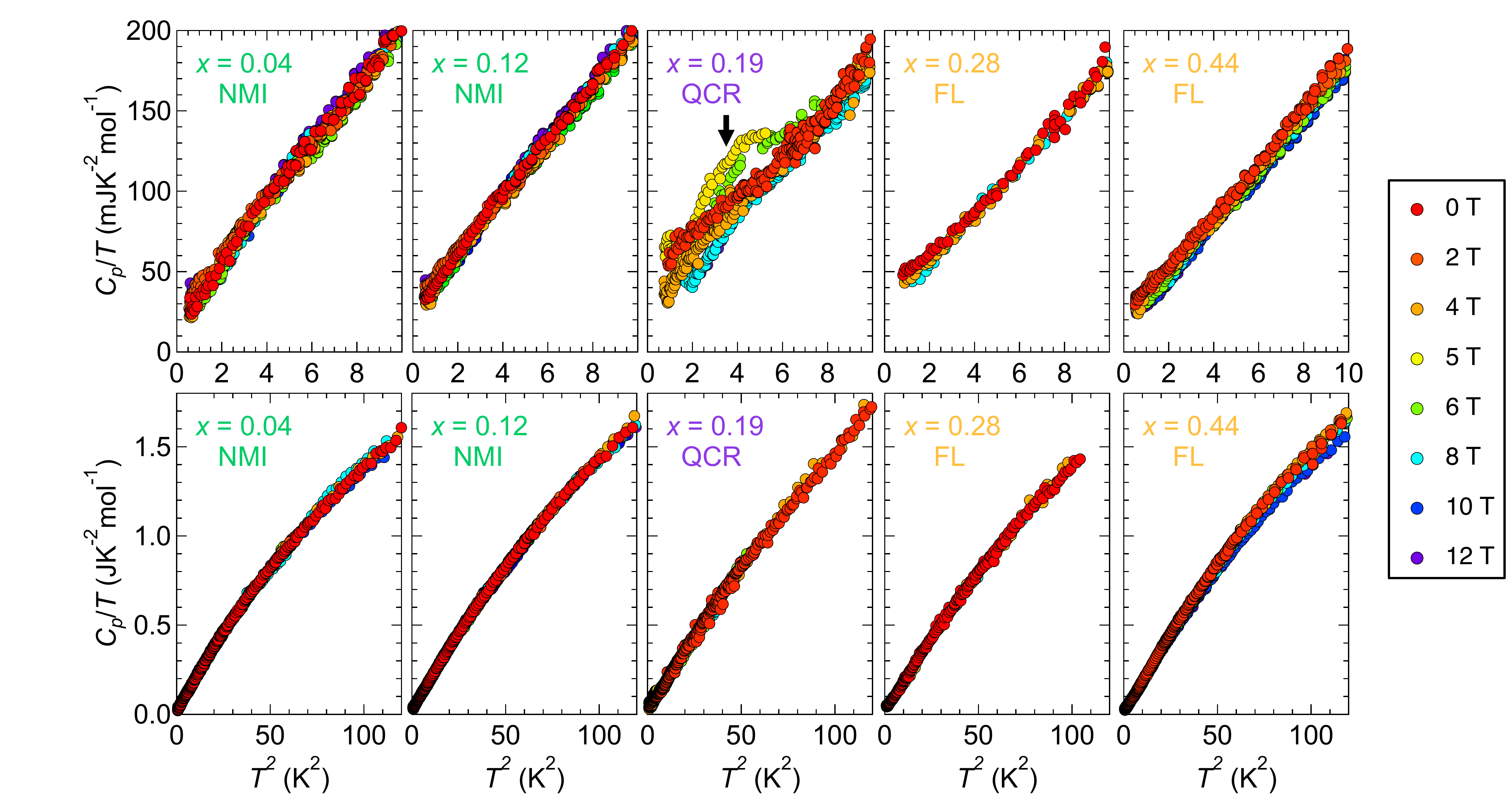}
\end{center}
\caption{(Color online)
$C_p$$T$$^{-1}$ vs. $T^2$ at various magnetic fields for $x$=0.04, 0.12 at the NMI, 0.19 at the quantum critical region (QCR), and 0.28, 0.44 at the FL.
The upper panels show the data below 10~K$^2$ while the lower ones show the data up to 120~K$^2$.
The arrow for the $x$=0.19 data indicates a hump observed at fields of 5-6~T.
}
\label{fig3}
\end{figure*}
%---------------------
The magnetic field dependences of  the heat capacity for  $x$=0.04, 0.12, 0.19, 0.28, and 0.44 are shown in Fig.~\ref{fig3}.
The upper panels show the low-temperature region below 10~K$^2$ while the lower ones display the data up to 120~K$^2$.
The fields are applied perpendicularly to the two-dimensional plane.
For the NMI ($x$=0.04) salt, the magnetic-field dependence is not significant even at high magnetic fields.
This fact indicates the robustness of these low-energy excitations against fields.
However, the response to the magnetic field for the $x$=0.19 sample, located in the quantum critical region (QCR), is distinct from those of the other salts.
The upturn observed at 0~T disappears with increasing magnetic field, while a broad hump structure in the temperature dependence of $C_p$$T$$^{-1}$ appears at relatively high magnetic fields of 5-6~T.
Since this behavior indicates that the low-temperature entropy shifts to higher temperatures in magnetic fields, the origin of this field dependence cannot be attributed to the 6~K anomaly and percolative superconductivity which is often observed near the 1st-order Mott transition.
Considering the magnetic field dependence of the Mott boundary\cite{36} and the bent quantum phase boundary\cite{30,31} (Fig.~\ref{fig1}(b)), the origin of the hump structure is also attributed to the critical behavior.
By further increasing fields, the broad hump is also suppressed, and the field dependence is diminished.
This behavior suggests that the high-field electronic state at low temperatures is out of the critical regime and can be regarded as the FL state.

\section{Discussion}
\begin{figure}
\begin{center}
\includegraphics[width=0.8\linewidth,clip]{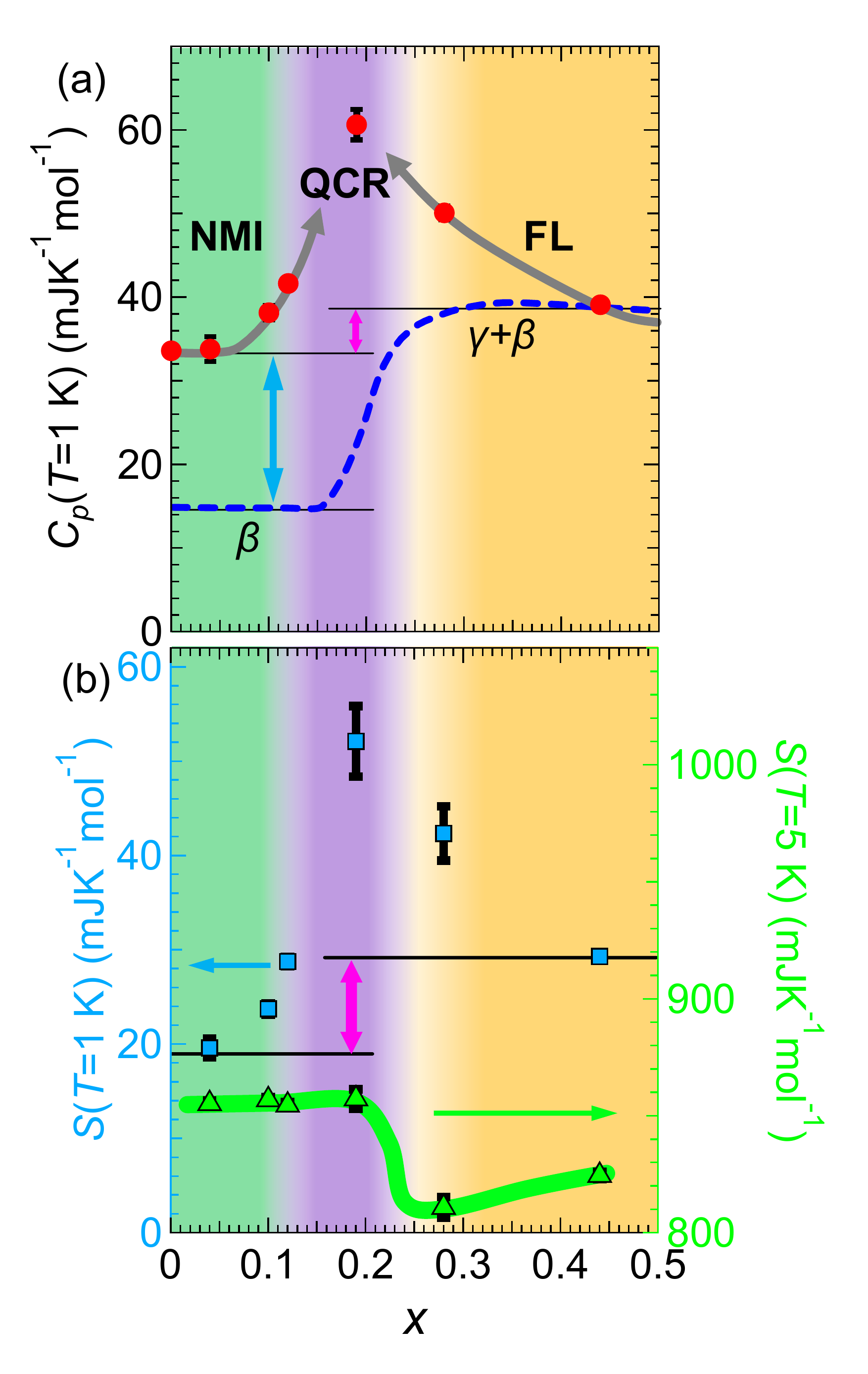}
\end{center}
\caption{(Color online)
Heat capacity $C_p$ at 1~K (a) and entropy $S$ at 1~K (left) and 5~K (right) (b) as a function of $x$.
The blue broken curve is the behavior expected based on the variation in $\gamma$ for the Mott transition between the AFI and FL states (Fig.~\ref{fig1}(c)).
The light blue and pink arrows in (a) highlight the contribution of the spin and charge sectors in the electronic heat capacity of the FL state, respectively.
The violet region represents the quantum critical region (QCR).
The thick translucent curve superimposed on the data points in (b) is a visual guide to make the $x$-dependence of $S$($T$=5~K) clearer.
}
\label{fig4}
\end{figure}

To deepen the understanding of the variations in the low-energy excitations around the Mott transition, we here show the low-temperature heat capacity at 1~K, $C_p$(1~K), as a function of $x$ in Fig.~\ref{fig4}(a).
%---------------------
%---------------------
In order to highlight the area near the Mott transition, each region is color-coded in a different color in the figure.
Based on the variation in $\gamma$ depending on $U$/$t$ (Fig.~\ref{fig1}(c)), $C_p$(1~K) should vary like the blue broken curve if the Mott transition is between the AFI and FL states.
Namely, the deviation from the blue broken curve is the peculiarity of the NMI-FL Mott transition.
For the $x$=0.04 salt, the value of 33.8~mJK$^{-2}$mol$^{-1}$, much larger than $\beta$=15~mJK$^{-4}$mol$^{-1}$ for the $x$=0.44 salt, suggests that the heat capacity involves finite low-energy excitations of the spin sector (the light blue arrow).
With approaching Mott transition, the $C_p$(1~K) increases from the constant value in the NMI region.
This behavior deviates from the blue broken curve because $\gamma$ is constantly zero inside the AFI Mott phase.
Once crossing the boundary and entering the FL regime, the $C_p$(1~K) asymmetrically decreases and reaches 39.1~mJK$^{-1}$mol$^{-1}$ at $x$=0.44, which is comparable with the typical value of $C_p$($T$=1~K)=$\gamma$+$\beta$ for the BEDT-TTF-based metallic salts with $\gamma$=20-25~mJK$^{-2}$mol$^{-1}$ and $\beta$=10-15 mJK$^{-4}$mol$^{-1}$.
The difference between the NMI and FL regions, shown by the pink arrow, should correspond to the contribution of the charge sectors of the $\pi$-electrons, which is absent in the NMI state.
If the inhomogeneity appearing near the first-order Mott transition develops around $x$=0.2, the strong enhancement of $C_p$(1~K) inside the FL region should not be observed near the boundary because the inhomogeneity significantly reduces the electronic heat capacity, as shown in Fig.~\ref{fig1}(c).

Here, we examine the slope of the phase boundary between the NMI and FL states on the electronic phase diagram, d$T$/d$x$ in Fig.~\ref{fig1}(d).
The positive slope indicates that the localization of electrons in the NMI state gives a larger entropy than the itinerancy of electrons in the FL state.
This unusual behavior is reminiscent of the Pomeranchuk effect observed in $^3$He, melting solid $^3$He with lowering the temperature through spin-lattice coupling\cite{37,38}.
In Fig.~\ref{fig4}(b), we present the $x$-dependence of the entropy $S$ at 1~K (left axis) and 5~K (right axis).
At 1~K, the entropy of the NMI state is lower than the entropy of the FL state.
At 5~K, it is the opposite.
As suggested by a theory\cite{36p5}, it is expected that there is only small energy difference between the FL and Mott states because the gain in kinetic energy of electrons in the FL state is compensated by the loss in potential energy.
In particular, when the frustration parameter $t$$^{\prime}$/$t$ is close to unity, the slope of the Mott boundary d$U$/d$T$ is almost zero or a small negative value\cite{36p6}, and thus, the energy difference between the two states should be very small.
This delicate energy balance makes the Mott transition winding on the phase diagram shown in Fig.~\ref{fig1}(c).
In real materials involving a variety of degrees of freedom, we must consider what contribution is an eventual factor determining how large or small the entropy is.
The similar $x$ dependence of $S$(1~K) and $C_p$(1~K) demonstrates that the low-temperature behavior can be explained by the electronic part and that the entropy of the NMI state is smaller at lower temperatures.
However, at higher temperatures above 5~K, the lattice part must also be considered because the lattice components account for a large portion ($>$90$\%$) of the total entropy in the soft organic crystal.
To explain the reversal of the entropy appearing with elevating temperature, we need to discuss entropy originating from the phonons as well as the low-energy spin excitations.
The characteristic of the present system is the confinement of the electrons in the triangular lattice making the antiferromagnetically interacting spins frustrated and disordered, which should result in lattice softening through the spin-lattice coupling.
The lattice softening entails shifting the phonon density of states down to a lower-temperature region, as is observed in the NMI state.
Low-energy phonon excitations in the non-ordered dimer-Mott triangular lattice system have been discussed by thermal conductivity measurements\cite{34p5}, and therefore, the softening of phonons can be a possible reason to explain the larger entropy in the NMI state.
Nevertheless, another possibility is that the spin excitations explain the evolution of entropy with temperature in the NMI state.
When d$U$/d$T$ is negative, the entropy of the NMI state can become larger than that of the FL state.
The formation of the possible non-magnetic VBS ordered state\cite{25,35} suggests a rapid increase in spin entropy with an enhancement of heat capacity near the transition temperature.
This gap-closing behavior around 5~K may also relate to the reversal of the entropy.
Although the spin entropy in the present system is not large, the relation may not be reversed even with the gain in the lattice entropy without the spin entropy.
For the NMI system, the low-temperature Pomeranchuk-like phase boundary\cite{38p5} is probably related to both contributions, namely the phonon softening effect and spin contributions.
To discuss these in more detail, the temperature dependence of entropy up to higher temperatures is necessary.

We emphasize that the low-temperature heat capacity $C_p$(1~K) should reflect the variation of the ground state driven by quantum fluctuations predominantly.
The gradual increase in $C_p$(1~K) with approaching the Mott boundary in the NMI phase suggests the continuous change in the low-energy excitations as the possible VBS state is suppressed near the Mott boundary.
According to the correspondence between the chemical pressure characterized by $x$ and physical pressure ($\Delta$$P$=1.5~kbar roughly corresponds to $\Delta$$x$=0.1)\cite{31} and the slope of the metal-insulator boundary d$x$/d$T$$\sim$2*10$^{-3}$~K$^{-1}$ (at $T$=5~K), the Clausius--Clapeyron relation d$P$/d$T$=$\Delta$$S$/$\Delta$$V$ leads to the volume change $\Delta$$V$$\sim$2*10$^{-8}$~m$^3$mol$^{-1}$ with the entropy difference $\Delta$$S$(5~K)$\sim$50~mJK$^{-1}$mol$^{-1}$.
Despite the rough estimation, the obtained $\Delta$$V$ is one order of magnitude smaller than the difference of $\Delta$$V$ between $x$=0 and 0.1, $\Delta$$V$$\sim$2*10$^{-7}$~m$^3$mol$^{-1}$\cite{26}.
Thus, even if the boundary is a first-order transition, its discontinuity must be almost negligible.
Near the QCP where the charge gap is just 0~K, quantum fluctuations related to the instability of the charge itinerancy are enhanced and destabilize the quasiparticles characterizing the FL.
It should be noticed that the quantum critical behavior is apparent only in the low-temperature region.
It is worthy to note that this energy scale is completely different from that of the high-temperature critical behavior induced by $U$ and $W$, which is commonly observed in all dimer-Mott systems irrespective of the geometrical frustration\cite{38p7}.
Indeed, the low-temperature critical behavior is absent in the less-frustrated system $\kappa$-(d[$n$,$n$]-BEDT-TTF)$_2$Cu[N(CN)$_2$]Br, which can access the 1st-order Mott transition between the AFI and the FL states\cite{36,36p1}.
As the peculiarity of the NMI salt is the persistence of the low-energy excitations related to the spin part, the present critical behavior may be induced by the instability of the fractionalization of the electron into the spin and charge sectors.
The present result and scenario agree with the discussion of the recent transport experiment under pressure\cite{9} and the thermodynamic investigation of $\kappa$-[(BEDSe-TTF)$_x$(BEDT-TTF)$_{1-x}$]$_2$Cu[N(CN)$_2$]Br\cite{39}, which is also another candidate hosting the genuine Mott transition between the NMI and FL state, as well as theoretical works\cite{23,24,39p5}.

Finally, we briefly discuss the so-called “6K anomaly” for the NMI sample, which has been discussed in the pristine $x$=0 sample\cite{25,29,40}.
Although the recent studies with high-quality samples and various sensitive measurements\cite{35,41} have allowed us to get closer to the details of this anomaly, the detail is still unclear because of some unresolved questions, such as the presence of the gapless excitations discussed above.
Since the pristine salts reported in the previous work\cite{29} are synthesized by other methods, their sample quality may differ from that of the present alloying series and quantitative comparison may be challenging.
Nevertheless, the systematic change in the physical parameters shown in Fig.~\ref{fig4} allows us to qualitatively compare our data with the results reported in the earlier work\cite{29}.
The data in Fig.~\ref{fig2}(e) indicates that the anomaly seems to be broadened and suppressed down to 3-4~K for the $x$=0.04 and 0.12 samples compared to that of the pristine sample.
The black dots shown in Fig.~\ref{fig1}(d) represent the $x$ dependence of the peak temperature of the anomaly.
Considering the relation to the charge disproportionation\cite{42}, it seems reasonable that the anomaly is smeared out by the suppression of the electron localization with approaching the Mott boundary.
In the lower panels in Fig.~\ref{fig3}, the magnetic field dependence of the anomaly is very small.
This feature robust against the magnetic field is consistent with the estimation of a critical field of order 60~T for the pristine salt\cite{25}.
To elucidate this enigmatic anomaly, more detailed investigations are desired in future studies.

\section{Conclusions}
In summary, we report the low-temperature thermodynamic properties for the chemical pressure tuning system of the dimer-Mott compounds that show no long-range ordering even at low temperatures.
The present result also provides evidence that the NMI state supports some gapless spin excitations.
However, we also found that this low-energy excitations do not seem to be described by a simple FL-like $\gamma$ term.
The systematic change in the heat capacity depending on $x$ revealed that the genuine Mott transition is potentially continuous via the QCP, which hosts the low-energy quantum fluctuations.
In the NMI state, the lattice softening originating from the geometrical frustrated lattice gives the larger heat capacity in total, although the opening of the charge gap reduces the electronic heat capacity.
Based on the entropy, the balance of these degrees of freedom makes the Pomeranchuk-like unique electronic phase diagram.
The coupling of these degrees of freedom makes the low-temperature phase competition between the NMI and FL states, leading to the low-energy quantum critical behavior that may be related to the instability of the fractionalization of the electron into the spin and charge sectors.

%acknowledgment
This work was partially supported by JSPS KAKENHI Grant No.19K22169 and 20H01862.

\end{document}